\documentclass[twocolumn,amsmath,amssymb,pra,showpacs]{revtex4}
\usepackage{setspace}
\usepackage{graphicx}
\usepackage{amsmath}
\usepackage{amssymb}
\usepackage{latexsym}

\usepackage{natbib}
\begin{document}

\title{Incoherence of Bose-Einstein condensates at supersonic speeds due to quantum noise}
\author{R.G. Scott$^{1}$, D.A.W. Hutchinson$^{2}$}
\affiliation{$^{1}$School of Physics and Astronomy, University of Nottingham, Nottingham, NG7 2RD, United Kingdom. \\$^{2}$The Jack Dodd Centre for Quantum Technology, Department of Physics, University of Otago, P.O. Box 56, Dunedin, New Zealand.}
\date{5/08/08}


\begin{abstract}
We calculate the effect of quantum noise in supersonic transport of Bose-Einstein condensates. When an obstacle obstructs the flow of atoms, quantum fluctuations cause atoms to be scattered incoherently into random directions. This suppresses the propagation of Cherenkov radiation, creating quantum turbulence and a crescent of incoherent atoms around the obstacle. We observe similar dynamics if the BEC is stirred by a laser beam: crescents of incoherent atoms are emitted from the laser's turning-points. Finally, we investigate supersonic flow through a disordered potential, and find that the quantum fluctuations generate an accumulation of incoherent atoms as the condensate enters the disorder. 
\end{abstract}

\maketitle

\section{Introduction}


The production of elementary excitations above a critical velocity is an ubiquitous feature of fluids. A familiar example is the sonic boom created as an aircraft exceeds the speed of sound in air. Recent experiments~\cite{Cornell} have demonstrated the analogous effect in dilute gas Bose-Einstein condensates (BECs): Cherenkov radiation which precedes an obstacle travelling at supersonic speeds relative to the BEC. This phenomenon has been reproduced theoretically~\cite{Cerenth,Cerenth2} by numerical integration of the Gross-Pitaevskii (GP) equation. However, unlike the aircraft, the BEC is a quantum object, so we might expect quantum fluctuations to play a significant role in the dynamics, especially because quantum noise has been shown to generate scattering halos and quantum turbulence for large collisional velocities~\cite{norrieprl,meotago}. Moreover, superfluids differ from normal fluids in that they have a critical velocity, known as the Landau critical velocity~\cite{pethick}, for the onset of dissipation. One of the early tests of superfluidity in BECs was to demonstrate a critical velocity for heating of the atom cloud~\cite{Ketcritv}. Although theoretical analysis of the GP equation has interpreted these results in terms of soliton and vortex production~\cite{Winiecki,Cerenth}, it is unclear whether quantum fluctuations are important. These ideas lead us to pose quite fundamental and general questions about how the coherence of BECs degrades during supersonic transport past an obstacle, a defect in the potential, or more extended potentials, either periodic or disordered. 

In this paper, we address these questions by performing hundreds of three-dimensional numerical simulations of dilute gas BECs using the Truncated Wigner method~\cite{steel,sinatra}. This allows us to map out the flow of coherent and incoherent particles as we vary the BEC velocity and the nature of the obstacle or defect. We begin by studying the supersonic transport of a BEC around an impenetrable cylindrical obstacle. Previous analysis of the GP equation has shown that a Mach cone develops behind (or in the wake of) the obstacle, containing solitons and vortices~\cite{Cerenth,Cerenth2}. In front of the obstacle, outside the Mach cone, a density modulation appears which is a form of Cherenkov radiation. Our calculations show that, in the presence of quantum fluctuations and strong inter-atomic interactions, this modulation exists only briefly, being almost completely destroyed at large times. Atoms are scattered incoherently into random directions, creating quantum turbulence~\cite{norrieprl} and a crescent of incoherent atoms in front of the obstacle. 

These findings lead us to re-visit the classic laser stirring experiments~\cite{Ketcritv}, which showed heating above a critical velocity approximately $25\%$ the bulk speed of sound. This effect has been attributed to vortex production~\cite{Winiecki,Cerenth}. Our simulations show that incoherent atoms may radiate from the oscillating laser but, in contrast to the vortex production, only at supersonic speeds since the Mach cone and Cherenkov radiation do not occur for subsonic transport. We quantify the incoherent scattering by plotting the peak incoherent density as a function of BEC velocity, and hence demonstrate the threshold behaviour at the bulk speed of sound. 

Finally, we explore the supersonic transport of BECs through a disordered potential, which is lower in amplitude than the BEC chemical potential. Our results add to the interpretation of recent experiments which studied damped dipole oscillations in disordered potentials~\cite{huletpra}. Despite the BEC velocity reaching deep into the supersonic regime in the experiments, hitherto no theoretical work has considered incoherent scattering. Theoretical work has been limited to studies of the GP equation, which has revealed coherent backscattering~\cite{backscat} and soliton production~\cite{albert}, suggesting that there may be significant heating and a reduction of the condensate fraction due to incoherent processes. Our simulations confirm this, revealing an accumulation of incoherent atoms as the condensate enters the disorder, and a corresponding suppression of the propagation of backscattered atoms. 

\section{System and methodology}

We consider a three-dimensional box, which has dimensions $L_{x}$, $L_{y}$ and $L_{z}$, filled with a $^{23}$Na BEC at constant density $n_{0} = 2.2 \times 10^{20}$ m$^{-3}$, with periodic boundary conditions. We expand the wavefunction $\Psi\left(x,y,z,t\right)$, where $t$ is time, over a basis of $M$ plane-wave states, which are propagated in time using the fourth-order Runga-Kutta in the interaction picture algorithm (RK4IP)~\cite{meotago,norrie,rob}. A flow is introduced by imposing a phase gradient in the $x$-direction. Due to the periodic boundary conditions, we are constrained by the condition that the phase must change by $2 \pi j$ over $L_{x}$, where $j$ is an integer. The velocity is therefore $v_{x} = \left(\hbar/m\right)\times\left(2\pi j/L_{x}\right)$. We investigate supersonic effects by increasing $v_{x}$ beyond the bulk speed of sound $v_{s} = \sqrt{h^{2}n_{0}a/\pi m^{2}}$, where $a=2.9$ nm is the s-wave scattering length and $m$ is the mass of a single $^{23}$Na atom. 

In order to investigate supersonic flow of BECs under different conditions, we introduce barriers and defects in the potential. We minimise non-equilibrium effects by varying the density across the barrier or defect such that the BEC chemical potential at $t=0$ is a constant throughout the grid. To avoid spurious effects associated with the periodic boundary conditions our simulations never run for longer than a maximum time $t_{\mbox{max}}=L_{x}/\left(2v_{x}\right)$.

Without quantum fluctuations, our scheme is equivalent to the projected GP equation~\cite{norrie}. The GP equation can describe the production of solitons, vortices, and Cherenkov radiation, but assumes complete coherence is always maintained. We include the effect of incoherent scattering in our model by adding random complex noise to each plane-wave mode. Heuristically, this simulates quantum vacuum fluctuations by adding classical fluctuations to the coherent field of the BEC's initial state. The amplitude of the noise has a Gaussian distribution, with an average value of half a particle per plane-wave mode, hence adding $\sim M/2$ ``virtual particles'' to the $\sim n_{0}L_{x}L_{y}L_{z}$ ``real particles''~\cite{foot3}. This technique is known as the Truncated Wigner method~\cite{steel,sinatra}. It models the dynamics of incoherently scattered atoms, which are forbidden within the GP formalism. Specifically, the technique has been applied to colliding BECs~\cite{norrieprl}, BEC reflection from a steep barrier~\cite{meotago}, three-body recombination processes~\cite{norriepra2}, collapsing BECs~\cite{collapse}, atom interferometers~\cite{scottjuddinterf}, BEC diffraction~\cite{scottjudddiff} and optical lattices~\cite{Janne}.

By running multiple simulations with different initial random quantum fluctuations, we may calculate the dynamics of coherent and incoherent atoms~\cite{norriepra}. Firstly, we compute the average density profile of the ensemble of simulations, then subtract the uniform density contribution $M/\left(2L_{x}L_{y}L_{z}\right)$ from the virtual particles. This is the density profile of \emph{all} atoms, coherent and incoherent. Secondly, we average $\Psi\left(x,y,z,t\right)$ over the ensemble of simulations, and compute the square modulus. This is the density profile of the \emph{coherent} atoms. We may then calculate the profile of \emph{incoherent} atoms by simply subtracting the coherent density profile from the total density profile.


\section{Flow past an inpenetrable obstacle}

\subsection{Single trajectory results}

\begin{figure}[tbp]
\centering
\includegraphics[width=1.0\columnwidth]{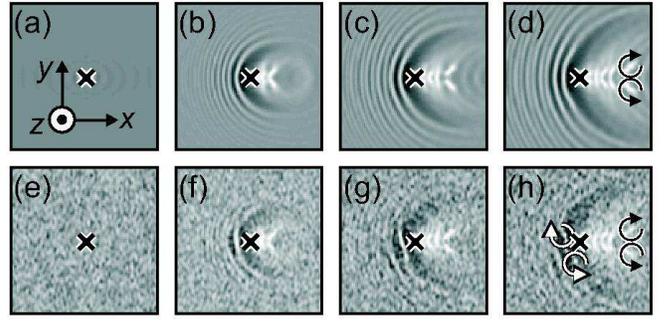}
\caption{Density profiles in the $z=0$ plane (axes inset) of single simulations of a BEC flowing at $v_{x}=1.5 v_{s}$ past an inpenetrable obstacle, whose position is indicated by a black cross, without (top row) and with (bottom row) quantum fluctuations, at $t=0$ [(a) and (e)], $t = 0.40$ ms [(b) and (f)], $t = 0.64$ ms [(c) and (g)], and $t = 0.88$ ms [(d) and (h)]. The field of view is $20\mu$m $\times$ $20\mu$m. Arrows in (d) and (h) indicate direction of quantized circulation around vortex cores.}
\label{f1}
\end{figure}

The top row of Fig.~\ref{f1} shows the well-known mean-field picture of Cherenkov radiation and vortex production as a BEC flows past an inpenetrable obstacle at a supersonic speed. These images are density profiles in the $z=0$ plane, calculated using the bare GP equation without quantum fluctuations. Hence, at $t=0$, the BEC has a uniform density away from the obstacle [Fig.~\ref{f1}(a)]. The obstacle, whose position is indicated by the black cross in Fig.~\ref{f1}(a), is a cylinder of diameter $1$ micron and length $L_{z}$, with its axis running parallel to the $z$-axis. Within the obstacle the atom density is set to zero. The imposed condensate velocity $v_{x} = 1.4 v_{s}$.

The Cherenkov radiation is observed as density modulations ahead of (to the left of) and curving around the obstacle [Fig.~\ref{f1}(b)-(d)], forming a cone shape known as the Mach cone. The angle $\theta$ between the flow direction and the Mach cone line is given by~\cite{Cerenth2}
\begin{equation}
\theta = \sin^{-1}\left(v_{s}/v_{x}\right) .
\label{coneangle}
\end{equation}
As $t$ increases from 0.40 [Fig.~\ref{f1}(b)] to 0.88 ms [Fig.~\ref{f1}(d)], the radiation gradually extends away from the obstacle. Vortices are seeded in the wake of (to the right of) the obstacle \emph{within} the Mach cone [Fig.~\ref{f1}(c) and (d)], and subsequently travel away in the direction of condensate flow. Two vortices are encircled by arrows in Fig.~\ref{f1}(d), indicating the direction of circulation.

The bottom row of Fig.~\ref{f1} shows the corresponding images including quantum fluctuations [Fig.~\ref{f1}(e)]. At $t=0.40$ ms [Fig.~\ref{f1}(f)], the Cherenkov radiation can be observed as it begins to form. However, it is not visible for $t \gtrsim 0.50$ ms [Fig.~\ref{f1}(g) and (h)], and has been replaced by a turbulent flow, including vortices [encircled by the white arrows in Fig.~\ref{f1}(h)] which are completely absent from the corresponding GP picture [Fig.~\ref{f1}(d)]. This turbulent flow has been referred to as ``quantum turbulence'' in the context of colliding BECs~\cite{norrieprl}. The vortices formed in the wake of (to the right of) the obstruction are unaffected by quantum fluctuations, and consequently appear in identical locations to those in the GP simulations. Two examples are encircled by the black arrows in Fig.~\ref{f1}(h), indicating the direction of circulation.

\subsection{Trajectory ensemble results}

\begin{figure}[tbp]
\centering
\includegraphics[width=1.0\columnwidth]{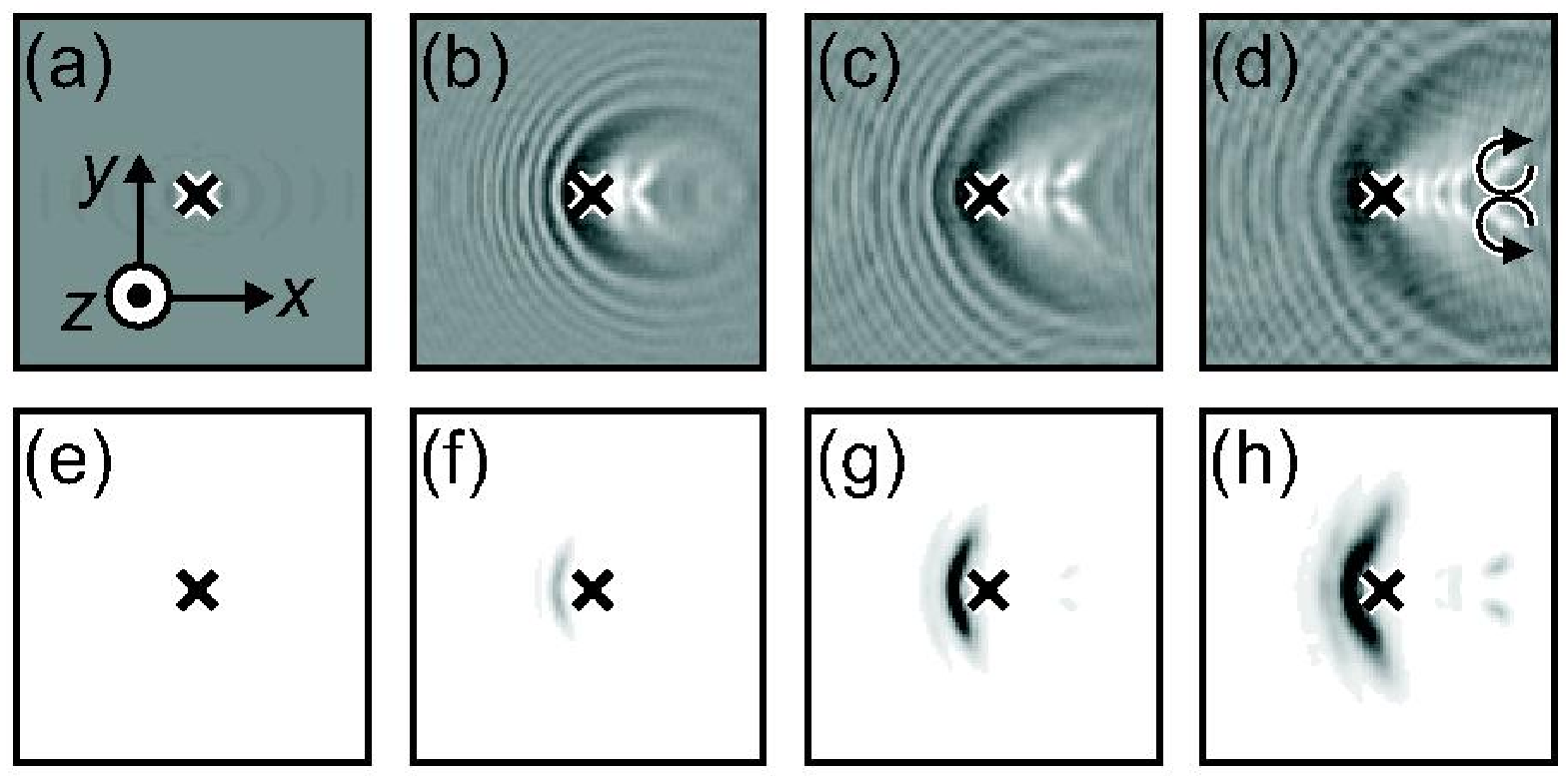}
\caption{Simulated absorption images in $x-y$ plane (axes inset) averaged over one hundred simulations of a BEC moving past an inpenetrable obstacle, whose position is indicated by a black cross, for $v_{x}=1.5 v_{s}$, at $t=0$ [(a) and (e)], $t = 0.40$ ms [(b) and (f)], $t = 0.64$ ms [(c) and (g)], and $t = 0.88$ ms [(d) and (h)]. Top (bottom) row shows total (incoherent) density. The field of view is $20\mu$m $\times$ $20\mu$m. Arrows in (d) indicate direction of quantized circulation around vortex cores.}
\label{f2}
\end{figure}

We may gain more insight by averaging multiple simulations with different initial quantum fluctuations. We calculate the average total density of one hundred simulations within a cube of dimensions $L_{x}=L_{y}=L_{z}=20$ $\mu$m, then integrate along the $z$-axis to obtain a two-dimensional density profile in the $x$-$y$ plane, equivalent to an experimental absorption image. The result is shown in the top row of Fig.~\ref{f2}. Since the quantum fluctuations are random, the initial profile of the averaged atom density appears uniform [Fig.~\ref{f2}(a)], except within the obstacle (whose position is indicated by the black cross) where the density is zero. These averaged images confirm that the Cherenkov radiation visible at $t=0.40$ ms [Fig.~\ref{f2}(b)] is strongly supressed by $t \gtrsim 0.50$ ms [Fig.~\ref{f2}(c) and (d)], in contrast to the clear density modulations found in the absence of quantum fluctuations [Fig.~\ref{f1}(c) and (d)]. As noted above, the production of vortices within the Mach cone [Fig.~\ref{f2}(b)-(d)] is identical to that in the simulations excluding quantum fluctuations [Fig.~\ref{f1}(b)-(d)]. 

The corresponding incoherent atom density profiles are shown in the lower row of Fig.~\ref{f2}. At $t=0$ the simulations are completely coherent, and hence the incoherent atom density profile is zero [Fig.~\ref{f2}(e)]. At $t = 0.40$ ms, when the Cherenkov radiation is still clearly visible [Fig.~\ref{f2}(b)], the incoherent density remains very low [Fig.~\ref{f2}(f)]. However, as the Cherenkov radiation begins to fade [Fig.~\ref{f2}(c)], a crescent-shaped concentration of incoherent atoms appears ahead of (to the left of) the obstruction [Fig.~\ref{f2}(g)]. The shape of the incoherent cloud reflects the regions of intense Cherenkov radiation. As $t$ increases to $0.88$ ms, the crescent of incoherent atoms gradually extends around the Mach cone and very slowly away from the cone [Fig.~\ref{f2}(h)], much slower than the Cherenkov radiation. These results show that the formation of Cherenkov radiation is suppressed by incoherent scattering in dense BECs with strong inter-atomic interactions, causing heating and a reduction in the condensate fraction. 

Cherenkov radiation can be distinguished from incoherently scattered atoms by studying their dynamics. As noted above, the center-of-mass of the incoherent atoms is approximately stationary (see Fig.~\ref{f2}). This is because these atoms have been scattered into random directions in the regions of intense Cherenkov radiation, and hence some atoms are travelling directly towards, or directly away from, the observer. In addition, multiple scattering events may occur, hence suppressing the propagation of the incoherent atoms. In contrast, the Cherenkov radiation is coherently emitted from the obstruction, travelling in a direction perpendicular to the surface of the Mach cone, as shown in Fig.~\ref{f1}. As an example, recent experiments~\cite{Cornell} have reported a shock-wave which propagates away from the obstacle during supersonic transport. This propagation identifies the shock-wave as Cherenkov radiation; incoherently scattered atoms would remain close to the obstacle. 

\begin{figure}[tbp]
\centering
\includegraphics[width=0.6\columnwidth]{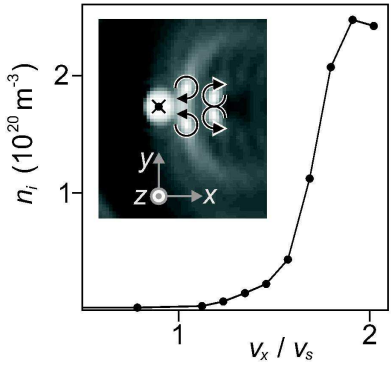}
\caption{$n_{i}$ versus $v_{x}$ at $t = 0.4 t_{\mbox{max}}$ for a BEC flowing past an inpenetrable obstacle. Inset shows the density profile in the $z=0$ plane (axes inset) for $v_{x}=0.8v_{s}$, without quantum fluctuations. The position of the obstacle is indicated by the black cross. The field of view is $10\mu$m $\times$ $12\mu$m. Arrows in inset indicate direction of quantized circulation around vortex cores.}
\label{f2b}
\end{figure}

In Fig.~\ref{f2b}, we quantify the incoherent scattering by plotting the averaged peak incoherent atom density $n_{i}$ as a function of $v_{x}$ at $t = 0.4 t_{\mbox{max}}$. The curve shows that $n_{i}$ rises abruptly from zero when $v_{x}$ exceeds $v_{s}$, approaching a maximum value, set by $n_{0}$, for $v_{x}\approx 2 v_{s}$. This threshold behavior occurs because the Mach cone and Cherenkov radiation do not form for flow speeds below the speed of sound~\cite{Cerenth2} ($\theta$ has a maximum value of $\pi/2$ for $v_{x}=v_{s}$, see Eq.~\ref{coneangle}). In contrast, vortex shedding can occur for flow speeds below the speed of sound, as has been observed experimentally~\cite{Ketcritv}. To illustrate this, the inset in Fig.~\ref{f2b} shows the density profile in the $z=0$ plane at $t=1.1$ ms ($=0.7 t_{\mbox{max}}$), for $v_{x}=0.8v_{s}$, without quantum fluctuations. No Mach cone or Cherenkov radiation is present, but two vortex anti-vortex pairs (encircled by arrows to indicate the direction of circulation) have been shed from the obstacle. Vortices can be generated for flow speeds below the speed of sound because the relative velocity between the obstacle and the BEC may exceed the speed of sound \emph{locally}~\cite{Ketcritv}. 

\section{Laser stirring}

The findings in the previous section motivated us to revisit the classic laser stirring experiment, in which heating was observed above a critical velocity~\cite{Ketcritv}. Theoretical work has attributed this heating to soliton and vortex production~\cite{Winiecki,Cerenth}, but incoherent scattering may play a significant role. We therefore replace the stationary obstacle in the previous section with an obstacle oscillating in the $x$-direction with speed $v_{o} = \pm1.8 v_{s}$ and amplitude $A = 4.0$ $\mu$m, and hence period $t_{o}=4A/v_{o}=1.2$ ms. The imposed phase gradient is also removed so the BEC is initially stationary.

\begin{figure}[tbp]
\centering
\includegraphics[width=1.0\columnwidth]{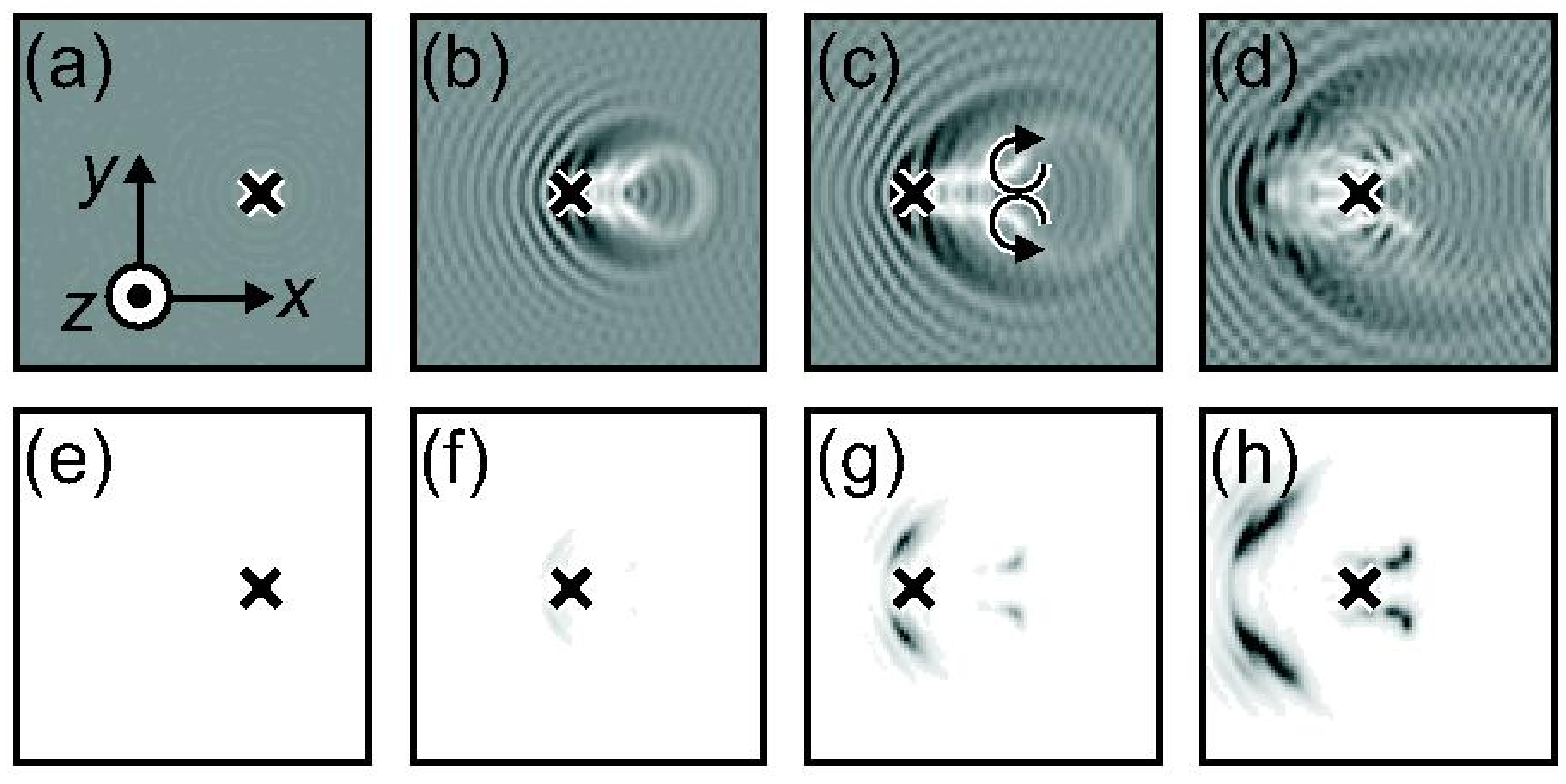}
\caption{Simulated absorption images in $x-y$ plane (axes inset) averaged over one hundred simulations of a BEC stirred by laser, whose position is indicated by a black cross, with $v_{o}=1.8 v_{s}$, at $t=0$ [(a) and (e)], $t = 0.36$ ms [(b) and (f)], $t = 0.58$ ms [(c) and (g)], and $t = 0.80$ ms [(d) and (h)]. Top (bottom) row shows total (incoherent) density. The field of view is $20\mu$m $\times$ $20\mu$m. Arrows in (c) indicate direction of quantized circulation around vortex cores.}
\label{f3}
\end{figure}

The top row of Fig.~\ref{f3} shows the simulated absorption images generated from one hundred laser drag simulations. The BEC density is uniform at $t=0$, except near the black cross which indicates the position of the laser [Fig.~\ref{f3}(a)]. Initially, the BEC is completely coherent, and hence the incoherent density is zero [Fig.~\ref{f3}(e)]. As the laser moves from right to left [Fig.~\ref{f3}(a)-(c)], Cherenkov radiation moves ahead of (to the left of) the laser, and vortices are generated in its wake (to its right). A crescent of incoherent atoms develops [Fig.~\ref{f3}(f) and (g)], moving from right to left ahead of the obstacle. We can again interpret this as incoherent scattering due to the Cherenkov radiation. 

For $t<t_{o}/2 = 0.58$ ms, the dynamics are identical to those produced by the static obstacle in the previous section: we have simply changed the observer's frame-of-reference. But after the laser reverses its direction of motion, the two systems are no longer equivalent. The Cherenkov radiation and the crescent of incoherent atoms continue to propagate in the original rest frame of the laser, whilst the laser travels in the opposite direction [Fig.~\ref{f3}(d) and (h)]. The supersonic motion of the Cherenkov radiation and incoherent atoms causes further incoherent scattering, broadening the incoherent region and raising the peak incoherent density. Meanwhile, the laser moves in the positive $x$-drection, and we now begin to observe the formation of a second crescent of incoherent atoms.

Incoherent scattering and subsequent quantum turbulence will contribute to the heating observed experimentally when a laser oscillates in a BEC~\cite{Ketcritv}. However, as shown in Fig.~\ref{f2b}, this additional heating occurs only when $v_{o} > v_{s}$, becoming significant for $v_{o} > 1.5 v_{s}$. The experiments focused instead on the threshold for heating, which occurs at a critical velocity of approximately $0.25 v_{s}$. In this threshold regime, the heating will be solely due to coherent vortex production~\cite{Winiecki,Cerenth}, because Cherenkov radiation does not occur below the bulk speed of sound. Consquently, these findings do not significantly change the interpretation of the experiments~\cite{Ketcritv}.

\section{Disordered potentials}

\begin{figure}[tbp]
\centering
\includegraphics[width=1.0\columnwidth]{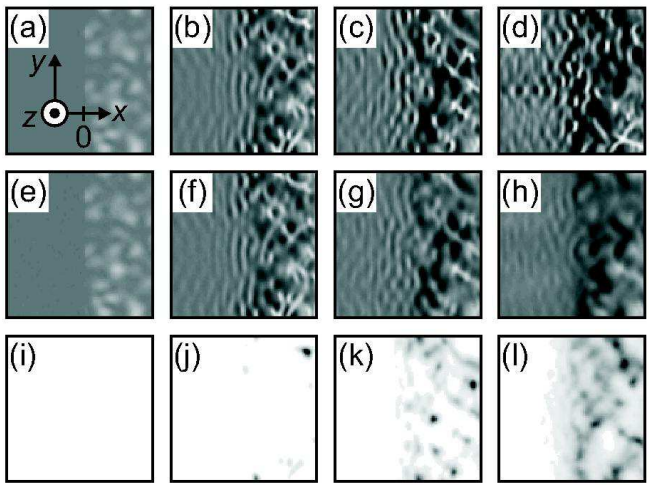}
\caption{Top row: simulated absorption images in $x-y$ plane (axes inset), without quantum fluctuations, of a BEC in a disordered potential, for $v_{x}=1.5 v_{s}$, at $t=0$ (a), $t = 1.0$ ms (b), $t = 2.0$ ms (c), and $t = 3.0$ ms (d). Middle (bottom) row: corresponding images including quantum fluctuations, averaged over one hundred simulations, showing total (incoherent) atom density. The field of view is $20\mu$m $\times$ $20\mu$m.}
\label{f5}
\end{figure}

The previous sections have discussed incoherent scattering from inpenetrable barriers, within which the potential exceeds the BEC chemical potential. However, significant scattering is also possible from small potential defects, lower in amplitude than the BEC chemical potential. To investigate this problem we generate a random potential with a peak amplitude of half the chemical potential. We set the potential to zero in the region $x<0$ in order to highlight the effect of backscattering from the random potential. The initial density of the BEC is adjusted so that its chemical potential is a constant. Consequently the simulated absorption image at $t=0$ is uniform for $x<0$ [Fig.~\ref{f5}(a), $x=0$ is indicated by the inset axes], and varies smoothly over the random potential for $x>0$. 

The top row in Fig.~\ref{f5} shows the dynamics of the BEC in the disordered potential without quantum fluctuations. The random potential causes a back-scattering~\cite{backscat}, which can be observed as the interference pattern close to the boundary between the flat and random potential regions at $x=0$ [Fig.~\ref{f5}(b)]. As time increases, this interference pattern increases in amplitude and extends away from the boundary in the negative $x$-direction [Fig.~\ref{f5}(c) and (d)]. 

If quantum fluctuations are included, the initial density profile is identical [Fig.~\ref{f5}(e)], and the corresponding incoherent density is zero [Fig.~\ref{f5}(i)]. Backscattering begins to occur [Fig.~\ref{f5}(f)] as before, but, rather than increasing in amplitude and extending away from the boundary region, the interference pattern begins to fade, and remains close to $x=0$ [Fig.~\ref{f5}(g) and (h)]. This is due to incoherent scattering between the coherently backscattered atoms and the incoming atoms approaching the disordered potential. Since the atoms are scattered into random directions, they remain close to the boundary region [Fig.~\ref{f5}(j)-(l)], rather than propagate away like the coherent backscattered atoms. This process introduces an additional contribution, which is not described within the GP equation, to the damping of BEC motion observed experimentally~\cite{huletpra}. There are also several regions of high incoherence \emph{within} the random potential, due to backscattering from local peaks in the potential. 

\section{Conclusions}

In summary, we have shown that supersonic transport of BECs can generate significant incoherent scattering. This causes a reduction of the condensed fraction, and hence an additional contribution to the loss of superfluidity and the supression of BEC motion, which is typically observed in experiment as heating and a damping of dipole oscillations~\cite{Ketcritv,huletpra}. We have mapped out the dynamics of the incoherent atoms by performing hundreds of three-dimensional simulations, and related our results to relevant experiments. 

In principle, our predictions can be tested experimentally, but in practice it may be difficult to separate heating due to coherent soliton or vortex production from heating due to incoherent scattering. It may be possible to exploit the drop in the condensed fraction as a result of incoherent scattering by, for example, applying a Bragg pulse to create fringes in the cloud. Due to loss of phase coherence, there will be a reduction in the fringe visibility~\cite{huletpra}. In contrast, coherent backscattering and vortex production will generate a reproducable fringe pattern. 


\begin{acknowledgments}
This work is funded by EPSRC and by the Government of New Zealand through the Foundation for Research, Science and Technology under New Economy Research Fund contract NERF-UOOX0703. 
\end{acknowledgments}

\bibliography{biblio}

\begin{thebibliography}{22}
\bibitem[{foo()}]{foot3}
\bibinfo{note}{Typically, the ratio of real particles to virtual particles is
  $\sim 50$:$1$.}
\end{thebibliography}

\end{document}